\documentclass[useAMS,usenatbib]{mn2e}
\usepackage{graphicx}

\title[High energy synchrotron emission]{On the mechanism of the pulsed high
energy emission from the pulsar PSR B1509-58}

\author[Chkheidze N. and Osmanov Z.]{Chkheidze N.\thanks{E-mail:
nino.chkheidze@iliauni.edu.ge} Osmanov Z. \\
Centre for Theoretical Astrophysics, ITP, Ilia State University, 0162-Tbilisi, Georgia\\
}
\begin{document}



\pagerange{\pageref{firstpage}--\pageref{lastpage}} \pubyear{2009}

\maketitle

\label{firstpage}

\begin{abstract}
We investigate the high-energy (HE) ($<1$GeV) emission from the
pulsar PSR B1509-58 and its relation to the radio emission in the
$1.4$GHz frequency band. The role of the quasi-linear diffusion in
producing the pulsed HE radiation is investigated. We show that by
means of the cyclotron instability the relatively low frequency
waves excite, which due to the diffusion process influence the
particle distribution function and switch on the synchrotron
emission mechanism. We argue that the coincidence of HE main peak
and the radio pulse is a direct consequence of the fact that the
high and low frequency radiation is produced simultaneously in a
local area of the pulsar magnetosphere. In the paper we also
consider the absence of the radio counter pulse and explain this
fact.
\end{abstract}


\section{Introduction}
Recent development in the field of high energy $\gamma$-ray
astronomy stimulates new theoretical approaches and methods in
modern astrophysics. Last decade the bulk of observational results
in the high and very high energy domains comes from several
telescopes: High Energy Stereoscopic System (HESS), Very Energetic
Radiation Imaging Telescope Array System (VERITAS), Major
Atmospheric Gamma-ray Imaging Cherenkov (MAGIC) Telescope etc. These
instruments have detected several sources at GeV and TeV energies
\citep{mariotti,abram,ongmar} and Fermi Large Area Telescope
(Fermi-LAT) in turn, announced the first catalogues of active
galactic nuclei (AGN) \citep{abdo1} and pulsars \citep{abdo2},
respectively.

In the context of the present paper a special interest deserves the
detection of the HE $<1$GeV $\gamma$-ray pulsed emission from the
young and energetic pulsar PSR B1509-58 by the Fermi-LAT with a
light curve presenting two symmetric peaks separated by phase
$0.37\pm0.02$ (see Fig.1 from \citet{abdo3}). The $\gamma$-ray
signals are offset from the main radio peak observed at $1.4$GHz by
phases $0.96\pm0.01$ and $0.33\pm0.02$. The first $\gamma$-ray peak
is almost coincident with the radio pulse. Particularly, leading the
main radio pulse by phase $0.04\pm0.01$. The results of Fermi-LAT
are obtained for the energy interval $30$MeV-$100$GeV and it is
shown that up to $1$GeV the emission reveals pulsations. We focus on
explaining the pulsed HE $\gamma$-ray emission of this source in
$1$MeV-$1$GeV interval.

According to the observational evidence, the main pulses in the
radio ($1.4$GHz) and HE ($<1$GeV) bands are almost coincident,
meaning that they are generated simultaneously, in a certain
localized area of the pulsar magnetosphere. In this context the
quasi-linear diffusion (QLD) could be an efficient mechanism
providing the coincidence of pulses in the aforementioned domains.
\citet{kmm} showed that if the magnetic energy density exceeds that
of plasmas the low frequency cyclotron modes are excited. In turn,
the unstable cyclotron waves interact with resonant particles by
means of the QLD and as a result they acquire the pitch angles,
leading to the efficient synchrotron emission process as it was
argued by \citet{lomin} and \citet{machus1}. For high
Lorentz-factors the synchrotron emission comes in the HE domain. In
a series of papers, the mechanism of QLD was applied to active
galactic nuclei \citep{difus3,difus4,difus5,difus6} and pulsars
\citep{malmach,nino,difus,difus1,ninoz}, respectively. Recently, we
have applied this mechanism to explain the pulsed very high energy
($>25$GeV) radiation of the Crab pulsar \citep{difus,difus1,ninoz}
and it's pulse-phase coincidence with the optical signals
\citep{magic}.

In the present paper we implement the mechanism of the QLD to the
pulsar PSR B1509-58. The observations confirm that the signals of
the HE $\gamma$-rays ($<1$GeV) are almost coincident in phase with
the radio signals ($1.4$GHz). Thus, we argue that they are produced
simultaneously in one location of the pulsar magnetosphere.

The paper is organized as follows. In Section 2 we consider our
model of the quasi-linear diffusion and apply it to the pulsar PSR
B1509-58, explaining major characteristics of emission; in Sect. 3
we discuss the model and in Sect. 4 we summarize our results.

\section{Model} \label{sec:consid}

According to the standard model of pulsars, it is believed that the
pulsar's magnetospheres is filled by electron-positron plasma, which
consists of three components. The so-called primary beam particles,
which uproot from the neutron star's surface and are described by
the Goldreich-Julian density \citep{GJ} in due course of time
accelerate in the gap electric field and reach a corresponding
threshold when the cascade pair creation becomes inevitable
\citep{stur,tadem}. Therefore, the pulsar's magnetosphere is filled
by the primary beam ($\gamma\sim\gamma_b$), the bulk of plasma
($\gamma\sim\gamma_p$) and the so-called tail particles
($\gamma\sim\gamma_t$) (see Fig1. from \citet{aro}). The
distribution function of the electron-positron plasma is
one-dimensional, as at the pulsar surface any transverse momenta of
relativistic electrons are lost in a very short time
(\(\leq10^{-20}\)s) via synchrotron emission in very strong magnetic
fields.

According to the work of \citet{kmm}, by means of the anomalous
Doppler effect plasmas in the pulsar magnetosphere undergo the
cyclotron instability. In particular, due to the curvature radius of
the field lines, $\rho$, the beam particles are forced to drift
perpendicularly to plane of the curved field line with the following
velocity
\begin{equation}\label{ux}
u_x\equiv \frac{cV_{_{_{\|}}}\gamma_b}{\rho\omega_B}
\end{equation}
where $c$ is the speed of light, $V_{_{\|}}$ is particle velocity
along the field line, $\omega_B\equiv eB/mc$ is the cyclotron
frequency, $B$ is the magnetic induction, $e$ and $m$ are the
electron's charge and the rest mass, respectively. The distribution
function is one-dimensional and anisotropic and plasma becomes
unstable. Both of these factors (the one-dimensionality of the
distribution function and the drift of particles) might cause
generation of eigen modes in the electron-positron plasma if the
following resonance condition is satisfied
\begin{equation}\label{eigen}
\omega - k_{_{\|}}V_{_{\|}}-k_xu_x-\frac{\omega_B}{\gamma_b} = 0.
\end{equation}
Here, $k_{_{\|}}$ is the longitudinal component of the wave vector
and $k_x$ is the wave vector's component along the drift. From Eq.
(\ref{eigen}) one can see that the pure transversal modes
($t$-waves) characterized by the following dispersion relation
\citep{kmm}.
\begin{equation}\label{omt}
\omega_t \approx kc\left(1-\delta\right),\;\;\;\;\; \delta =
\frac{\omega_p^2}{4\omega_B^2\gamma_p^3},
\end{equation}
where $k$ denotes the modulus of the wave vector, $\omega_p \equiv
\sqrt{4\pi n_pe^2/m}$ is the plasma frequency and $n_p$ is the
plasma density. In the framework of the present model, the
aforementioned waves, by means of the QLD, influence the resonant
particles leading to the diffusion of particles along and across the
magnetic field lines. Consequently, particles acquire pitch angles
and start to radiate in the synchrotron regime.

It should be mentioned that we study the process of the QLD
close to the light cylinder surface, where, as it was investigated,
effects of centrifugal acceleration are very efficient for pulsars
\citep{or09} and AGN \citep{osm7}. Particularly, for explaining high
and very high energy emission of pulsars one needs a mechanism
providing very high energies of electrons, which is not provided by
the gap acceleration process. As it was shown in \citet{or09}, the centrifugal acceleration process
is quite efficient in the light cylinder zone and can
provide the reacceleration of the primary beam particles to reach
the Lorentz factors of the order of $10^7$. This in turn leads to the HE emission
pattern, that is in a good agreement with the observational data
\citep{or09}.

In order to describe the process of the QLD, one has to take into
account as the dissipative factors, which try to decrease the pitch
angles, as the diffusion attempting to increase them. When
relativistic electrons emit in the synchrotron regime, they undergo
the so-called reaction force, having as longitudinal as transversal
components  \citep{landau}
\begin{equation}\label{f}
    F_{\perp}=-\alpha\frac{p_{\perp}}{p_{\parallel}}\left(1+\frac{p_{\perp}^{2}}{m^{2}c^{2}}\right),
    F_{\parallel}=-\frac{\alpha}{m^{2}c^{2}}p_{\perp}^{2},
\end{equation}
where $\alpha=2e^{2}\omega_{B}^{2}/3c^{2}$, $p_{\perp}$ and
$p_{\parallel}$ are transversal and longitudinal components of
momentum, respectively.

On the other hand, the diffusion process, under certain conditions,
might balance the corresponding dissipation, leading to  the
quasi-stationary state ($\partial /\partial t = 0$), which, as is
shown by Chkheidze et at. (2011), for $\partial/\partial
p_{\perp}>>\partial/\partial p_{\parallel}$ results in the following
kinetic equation governing the aforementioned mechanism
\begin{eqnarray} \label{qld}
    \frac{\partial}{\partial p_{\perp}}\left(p_{\perp}
    F_{\perp}\textit{f }\right)=\frac{\partial}{\partial p_{\perp}}\left(p_{\perp}
D_{\perp,\perp}\frac{\partial\textit{f }}{\partial
p_{\perp}}\right),
\end{eqnarray}
where $\textit{f }$ is the distribution function of electrons,
\begin{eqnarray} \label{Dpp}
D_{\perp,\perp}\equiv\frac{e^2|E_k|^2\delta }{8c},
\end{eqnarray}
is the diffusion coefficient and $E_k$ is the electric field. In
order to estimate the value of $|E_k|^2$, we assume that half of the
beam energy density, $mc^2n_b\gamma_b/2$ ($n_b$ is the beam
density), converts to the energy density of the waves $|E_k|^2k$.
Therefore, an expression of $|E_k|^2$ writes as follows
\begin{equation}\label{ek2}
|E_k|^2 = \frac{mc^3n_b\gamma_b} {2\omega}.
\end{equation}

The solution of Eq. (\ref{qld}) is given by
\begin{equation}\label{ff}
    \textit{f}(p_{\perp})=C exp\left(\int \frac{F_{\perp}}{D_{\perp,\perp}}dp_{\perp}\right)=Ce^{-\left(\frac{p_{\perp}}{p_{\perp_{0}}}\right)^{4}},
\end{equation}
where
\begin{equation}\label{pp0}
     p_{\perp_{0}}=\left(\frac{4\gamma_bm^3c^3D_{\perp,\perp}}{\alpha}\right)^{1/4},
\end{equation}
and for the mean value of the pitch angle we obtain
\begin{equation}\label{pitch}
\langle\psi\rangle
 = \frac{p_{\perp_{0}}}{2\gamma_b mc}.
\end{equation}

\subsection{Radio emission}

According to the present model, the generation of the synchrotron
emission is provided by the feedback of the low frequency cyclotron
waves on the resonant particles. It is easy to show from Eqs.
(\ref{eigen}, \ref{omt}) that the unstable cyclotron waves will have
the following frequency \citep{machus1}
\begin{equation}\label{nu}
\nu\simeq
\frac{e^{2}B^{2}P}{2\pi^{2}m^{2}c^{2}}\frac{\gamma_{p}^{2}}{\gamma_{b}^{4}}\times10^{-9}
GHz.
\end{equation}
The estimations show that for reasonable magnetospheric parameters
(we assumed that $\gamma_{p}\sim1$ and $\gamma_{b}\sim10^{7}$
\citep{or09}) the anomalous Doppler effect leads to the excitation
of the radio waves ($1.4$GHz), which in turn influences the
particles and switches on the synchrotron radiation process giving
the HE $\gamma$-ray emission. According to \cite{abdo3} it is shown
that the main pulses of the HE ($<1$GeV) $\gamma$-rays and the radio
emission come with the same phases. We argue that this observation
feature is caused by the fact that the emission in both domains are
produced simultaneously and in one location of the magnetosphere -
close to the light cylinder surface.

\subsection{Spectrum of the high energy emission}

In this subsection we are going to explain the observed spectrum of
the HE $\gamma$-rays. Following the method developed by Ginzburg
(1981), one can show that the flux of the synchrotron emission is
given by the following expression \citep{ninoz}
\begin{equation}\label{flux}
    F_{\epsilon}\propto\int_{p_{_{\parallel _{min}}}}^{p_{_{\parallel _{max}}}}\textit{f}_{_{\parallel}}(p_{_{\parallel}})B\langle\psi\rangle\frac{\epsilon}{\epsilon_{_{GeV}}}\left[\int_{\epsilon/
    \epsilon_{_{GeV}}}^{\infty}K_{5/3}(z)dz \right] dp_{_{\parallel}},
\end{equation}
where $\textit{f}_{_{\parallel}}(p_{_{\parallel}})$ is the
longitudinal distribution function of electrons. The major
difference from the model of Ginzburg is that according to our
scenario the synchrotron emission is generated in the outer part of
the pulsar magnetosphere via the QLD, which inevitably restricts the
values of the pitch angles. On the other hand, according to Ginzburg
(1981), along the line of sight the magnetic field is chaotic
leading to the broad interval of the latter $\psi \in(0; \pi)$. This
difference is principle between the standard approach and the method
presented in this paper. Therefore, in Eq. (\ref{flux}) instead of
$\psi$ we put it's average value (see Eq. (\ref{pitch})), which,
unlike the standard model, is the dynamical parameter and
participates in integration.

In order to find out the spectrum of the HE emission, one needs to
know the behaviour of the distribution function,
$\textit{f}_{_{\parallel}}(p_{_{\parallel}})$. For solving this
problem, we consider the equation governing the evolution of
$\textit{f}_{_{\parallel}}(p_{_{\parallel}})$ \citep{ninoz}
\begin{equation}\label{fpar}
    \frac{\partial\textit{f}_{_{\parallel}}}{\partial t}=\frac{\partial}{\partial
    p_{_{\parallel}}}\left({\frac{\alpha_{s}}{m^{2}c^{2}\pi^{1/2}}p_{\perp_{0}}^{2}\textit{f}_{_{\parallel}}}\right).
\end{equation}
We are interested in the stationary state ($\partial/\partial t =
0$), of the QLD, which reduces the aforementioned equation and leads
to the following form of the distribution function
\begin{equation}\label{fpar1}
    \textit{f}_{_{\parallel}}\propto\frac{1}{p_{_{\parallel}}^{1/2}|E_{k}|}.
\end{equation}

On the other hand, the cyclotron noise is described by the equation
\begin{equation}\label{noise}
    \frac{\partial|E_{k}|^{2}}{\partial
    t}=2\Gamma|E_{k}|^{2}\textit{f}_{_{\parallel}},
\end{equation}
where the increment of the instability, $\Gamma$, is given by
\citep{kmm}
\begin{equation}\label{gam}
   \Gamma=\pi \frac{\gamma_b\omega_b^2\delta}{\gamma_p\omega_B}.
\end{equation}
Here $\omega_b\equiv\sqrt{4\pi n_b e^2/m}$ is the plasma frequency
of the resonant (beam) electrons.

By taking into account Eqs. (\ref{noise},\ref{gam}) one can reduce
Eq. (\ref{fpar}) to the following form \citep{ninoz}
\begin{equation}\label{fpar2}
   \textit{f}_{\parallel}-\beta\frac{\partial}{\partial
    p_{_{\parallel}}}\left(\frac{|E_{k}|}{p_{_{\parallel}}^{1/2}}\right)=\textit{f}_{_{\parallel {0}}},
\end{equation}
where
\begin{equation}\label{beta}
   \beta =\left(\frac{4}{3}\frac{e^{2}}{\pi^{5}c^{5}}\frac{\omega_{B}^{6}\gamma_{p}^{3}}{\omega_{p}^{2}}\right)^{1/4},
\end{equation}
and $\textit{f}_{_{\parallel {0}}}$ is the initial distribution
function of particles. Since the electron number density behaves
with distance as $1/r^3$, it is clear that one has the following
condition, $\textit{f}_{_{\parallel }}\ll\textit{f}_{_{\parallel
{0}}}$, that reduces Eq. (\ref{fpar2})
\begin{equation}\label{fpar3}
   \beta\frac{\partial}{\partial
    p_{_{\parallel}}}\left(\frac{|E_{k}|}{p_{_{\parallel}}^{1/2}}\right)+\textit{f}_{_{\parallel {0}}}=0.
\end{equation}

\begin{figure}
  \centering {\includegraphics[width=7cm]{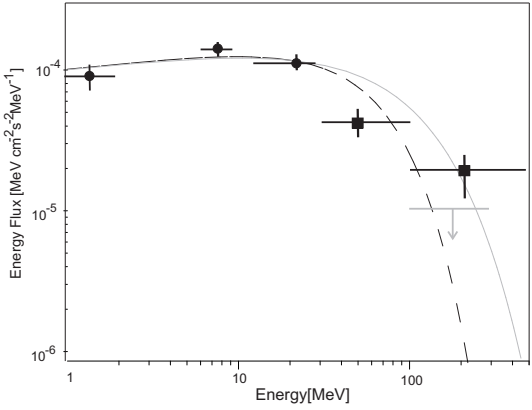}}
  \caption{Spectral energy distribution of PSR B1509-58 (dashed-line corresponds to the model spectrum $\epsilon^{-1.87}exp[-(\epsilon/0.059)^{1.3}]$
  and the solid-line to
  $\epsilon^{-1.87}exp[-(\epsilon/0.081)]$ obtained by \citet{pilia} from a fit of pulsed fluxes from soft to hard $\gamma$-rays.
  The round points represent COMPTEL and square points AGILE
  observations \citep{kupi,pilia}, respectively. The arrow represents Fermi upper
  limit \citep{abdo3}.
}\label{fig1}
\end{figure}

As we can see the function $E_{k}(p_{\parallel})$ drastically
depends on the form of the initial distribution of the primary beam
electrons.  According to the work, \citep{GJ}, a spinning magnetized
neutron star generates an electric field which extracts electrons
from the star's surface and accelerates them to form a low-density
($n_{b}=B/Pce$) and energetic primary beam, which in turn, is
reaccelerated in the light cylinder zone and might reach quite high
Lorentz factors, $\sim 10^7$, \citep{or09}. We only know the
scenario of creation of the primary beam, but nothing can be told
about its distribution, which drastically depends on the neutron
star's surface properties and temperature. To our knowledge there is
no convincing theory which would predict the initial form of the
distribution function of the beam electrons. Thus, we can only
assume that the beam electrons have the power-law distribution,
$\textit{f}_{_{\parallel {0}}}\propto p_{_{\parallel}}^{-n}$, and
for $|E_{k}|^2$ we will obtain the following behaviour
\begin{equation}\label{ekk}
     |E_{k}|^{2}\propto p_{_{\parallel}}^{3-2n}.
\end{equation}
In order to simplify Eq. (\ref{flux}), we replace the integration
variable $p_{_{\parallel}}$ by
$x\equiv\epsilon/\epsilon_{_{GeV}}\simeq3.6\times10^{18}\nu\gamma^{-9/4}$,
and taking into account Eqs.
(\ref{Dpp},\ref{pp0},\ref{pitch},\ref{fpar1},\ref{ekk}) the
expression of the flux will be
\begin{equation}\label{flux1}
    F_{\epsilon}\propto\epsilon^{-\frac{2-n}{4-n}}\int_{x_{min}}^{x_{max}} x^{\frac{2-n}{4-n}}\left[\int_{x}^{\infty}K_{5/3}(z)dz \right] dx.
\end{equation}
As we see from this equation the HE spectral index is given by the
following expression $(2-n)/(4-n)$. According to \citet{pilia} the
observed HE pulsed emission of PSR B1509-58 in the energy domain
($0.001-1$)GeV is best described by a power-law plus cutoff with the
spectral index equal to $1.87$. When $n=6.3$ the spectral index of
the synchrotron emission equals $1.87$ {\bf (see Eq.
(\ref{flux1}))}. If we assume that the energy of the emitting
electrons vary between $\gamma_{min}\simeq3\cdot10^{6}$ and
$\gamma_{max}\simeq3.5\cdot10^{7}$, one can show that the integral
(\ref{flux1}) can be approximately expressed by the following
function
\begin{equation}\label{spectrum}
    F_{\epsilon}\propto\epsilon^{-1.87}exp\left[-\left(\frac{\epsilon}{0.059}\right)^{1.3}\right].
\end{equation}
As we can see our emission scenario predicts the exponential cutoff,
with the cutoff energy $59$MeV (see Fig.~\ref{fig1}).

\section{Discussion}
It is clear that the present model provides the self consistent
explanation of several interesting observational features of the
pulsar PSR B1509-58. In particular, we have already seen that for
reasonable magnetospheric parameters the QLD can support the HE
pulsed emission up to $1$GeV. According to the observations, as it
was shown by \citet{abdo1}, the main pulses of the HE domain are
almost coincident with those of the radio emission on the frequency
$1.4$GHz. In the framework of our model this might mean that
radiation in both domains are produced almost simultaneously and in
one location of the pulsar magnetosphere. We have shown that by
means of the cyclotron instability, the radio waves are excited,
which influence the resonant particles and cause their diffusion as
along also across the magnetic field lines. Consequently, they
acquire the pitch angles and the synchrotron emission process is
switched on. This is the reason, why the HE $\gamma$-rays and the
low frequency waves are strongly connected. On the other hand we
neglect the role of the curvature radiation, because otherwise the
main pulse of the HE emission would be wider than it is observed.
This automatically means that the curvature radius of the magnetic
field lines are large enough to insure the negligible role of the
curvature radiation process. As it was shown by Osmanov et al.
(2008,2009) such a configuration of the field lines might be
provided by means of the so-called curvature drift instability, that
makes the magnetic field lines rectify efficiently in the very
vicinity of the light cylinder surface. In particular, as it was
shown in the aforementioned articles, the curvature drift (see Eq.
(\ref{ux})) will create the corresponding current, leading to the
creation of additional magnetic field. On the other hand, this
toroidal field, by means of the centrifugal effects, will
efficiently amplify, changing the overall configuration of the
magnetosphere. It has been found that the curvature drift
instability has two general modes, one of which makes the field
lines more curved and the other makes them rectify. We assume that
since the HE profile of the main signal is relatively narrow, the
curvature radiation should be negligible and therefore, the field
lines must be almost straight, which is provided by the curvature
drift instability.

In the framework of the present paper we provide theoretical
confirmation of the measured HE spectrum of PSR B1509-58 in the
energy domain $(0.001-1)$GeV. In particular, our model predicts the
power-law plus an exponential cutoff, which provides an explanation
of the observed decrease of the flux above $10$MeV \citep{abdo3}. If
the Lorentz factor of the resonant beam electrons are
$\gamma_{b}\simeq3\cdot10^{6}-3.5\cdot10^{7}$, then our theoretical
spectrum
$F_{\epsilon}\propto\epsilon^{-1.87}exp[-(\epsilon/0.059)^{1.3}]$.
As we can see the model automatically yields the exponential cutoff,
with the cutoff energy $59$MeV.

Another important feature of the high energy $\gamma$-rays is that
unlike the radio emission, apart from the main pulse it also has the
second pulse, separated from it by the phase $\Delta\phi = 0.37\pm
0.02$.

\begin{figure}
  \centering {\includegraphics[width=9cm]{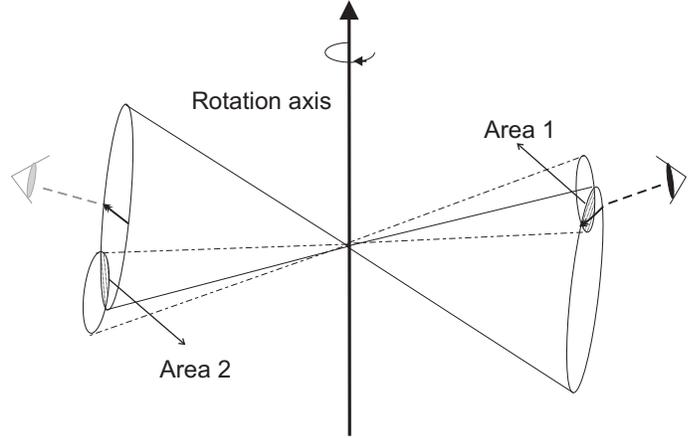}}
 \caption{Emission geometry of PSR B1509-58.
}\label{fig2}
\end{figure}

The observationally evident fact that the HE emission reveals two
peaks might be a direct result of an emission geometry. In
particular, we detect the pulsar emission twice per one rotation if
the HE emission cone is almost perpendicularly inclined with respect
to the rotation axis. According to the observations the main
$\gamma$-ray pulse leads the radio pulse by phase $0.04$, which
indicates that the emission cones of the HE and radio domains are
not spacially coincident (see Fig.~\ref{fig2}). In Fig.~\ref{fig2}
cone 1 represents the HE emission cone, that has a common area (area
1 and area 2) with the radio emission cone (cone 2). Then it is
clear that if the emission comes from area 1 which is parallel to
the line of sight, we will see the emission profiles for the radio
and HE bands, respectively. Due to star rotation and the
corresponding emission channels, the counter HE radiation cone will
have a position convenient for detecting the counter signal, whereas
the counter radio cone will be significantly shifted, and the line
of sight will be out of it. Therefore, as a result, we see the
signal and the counter signal of HE emission and do not see the
counter radio peak.

This approach explains another feature of the HE emission as well.
As it is evident from the observations, the counter peak is wider
with respect to the main pulse. In particular, from Fig.~\ref{fig2}.
it is clear that the time during which the line of sight stays
inside the cone 1 is shorter than the time scale it stays inside
counter HE cone.

\section*{Acknowledgments}
We thank Prof. George Machabeli for valuable discussions.

\label{lastpage}

\end{document}